\RequirePackage{lineno}
\documentclass[aps,prl,showpacs,twocolumn,groupedaddress]{revtex4}  

\usepackage{graphicx}  
\usepackage{dcolumn}   
\usepackage{bm}        
\usepackage{amssymb}   
\usepackage{epsfig}
\RequirePackage[usenames]{pstcol}

\def\met{{\mbox{$E\kern-0.57em\raise0.19ex\hbox{/}_{T}$}}}

\def\lmet{$WH\rightarrow \ell\kern-0.45em\raise0.19ex\hbox{/} \nu b\bar{b}$}

\def\hbb{$H\rightarrow b\bar{b}$}

\newcommand{\ttbar}{$t\overline{t}$}

\newcommand{\MET}{$\not\!\!E_T$}

\begin{document}

\leftline{.}
\rightline{FERMILAB-PUB-08-297-E}









\title{A search for associated $W$ and Higgs Boson production  in $p \bar{p}$ collisions at $\sqrt{s}=1.96$~TeV}

%
\author{V.M.~Abazov$^{36}$}
\author{B.~Abbott$^{75}$}
\author{M.~Abolins$^{65}$}
\author{B.S.~Acharya$^{29}$}
\author{M.~Adams$^{51}$}
\author{T.~Adams$^{49}$}
\author{E.~Aguilo$^{6}$}
\author{M.~Ahsan$^{59}$}
\author{G.D.~Alexeev$^{36}$}
\author{G.~Alkhazov$^{40}$}
% ...

\affiliation{\vspace{0.1 in}(The D\O\ Collaboration)\vspace{0.1 in}}
% ...
\affiliation{$^{6}$University of Alberta, Edmonton, Alberta, Canada,
                Simon Fraser University, Burnaby, British Columbia, Canada,
                York University, Toronto, Ontario, Canada, and
                McGill University, Montreal, Quebec, Canada}
% ...
\affiliation{$^{29}$Tata Institute of Fundamental Research, Mumbai, India}
% ...
\affiliation{$^{36}$Joint Institute for Nuclear Research, Dubna, Russia}
% ...
\affiliation{$^{40}$Petersburg Nuclear Physics Institute,
                St. Petersburg, Russia}
% ...
\affiliation{$^{49}$Florida State University, Tallahassee, Florida 32306, USA}
% ...
\affiliation{$^{51}$University of Illinois at Chicago,
                Chicago, Illinois 60607, USA}
% ...
\affiliation{$^{59}$Kansas State University, Manhattan, Kansas 66506, USA}
% ...
\affiliation{$^{65}$Michigan State University,
                East Lansing, Michigan 48824, USA}
% ...
\affiliation{$^{75}$University of Oklahoma, Norman, Oklahoma 73019, USA}
% ...

\date{August 14$^{th}$ 2008}

\begin{abstract}
We present results of a  search for $WH \rightarrow \ell \nu b \bar{b}$ production
in $p\bar{p}$ collisions
based on the analysis of 1.05 fb$^{-1}$ of data
collected by the D0 experiment at the Fermilab Tevatron, using
a neural network for separating the signal from backgrounds.
No signal-like excess is observed, and we set 95\% C.L.
upper limits on the {\it WH} production cross section multiplied by the branching ratio
for \hbb\ for Higgs boson masses between 100 and 150~GeV.
For 
a 
mass of 115 GeV,
we obtain an observed (expected) limit of 1.5~(1.4)~pb, a factor of 
11.4 (10.7)  times larger than
standard model prediction.
\end{abstract}
\maketitle


The Higgs boson is the last  unobserved particle of the standard model (SM).
 As a remnant
of spontaneous electroweak symmetry breaking, it is
fundamentally different from the other elementary particles, and its observation
would support the hypothesis that the Higgs mechanism  generates
the masses of the weak gauge bosons and the charged fermions. The Higgs boson
mass ($m_H$) is not theoretically predicted, but
the combination of results from direct searches at the CERN LEP 
collider~\cite{sm-lep}
with the indirect constraints from precision electroweak measurements 
results in a preferred range  of
$ 114.4<m_H<190$  GeV at 95\% C.L~\cite{elweak}. 
Such mass range can be probed at the Fermilab Tevatron collider.
%
In this Letter, we concentrate on the most sensitive production channel at the Tevatron
for Higgs bosons of mass below 125 GeV, i.e. the associated production of a Higgs boson
with a $W$ boson.
%
Several searches for {\it WH} production
 have  been  published
at a center-of-mass energy of $\sqrt{s}=1.96\,\mbox{TeV}$.
Two~\cite{emu-hep-ex/0410062,wh-plb}
used subsamples (0.17~fb$^{-1}$ and 0.44~fb$^{-1}$)
of the data reported in this Letter, while
two others,  from the CDF collaboration, are based on
0.32~fb$^{-1}$ and
0.95~fb$^{-1}$ of integrated luminosity~\cite{emu-CDF-wh,emu-CDF-wh-1fb}.

This analysis uses 1.05 fb$^{-1}$ of D0~\cite{emu-run1det,emu-run2det} data,
collected between April 2002 and February 2006.
As in our previous {\it WH} analyses~\cite{emu-hep-ex/0410062,wh-plb},
we require one high transverse momentum ($p_T$) lepton ($e$ or $\mu$) and
missing transverse 
energy~\MET\ to account for the neutrino from~the $W$ boson~decay,
 and two jets from the decay of the                               
 Higgs boson, with at least one of them
 being identified as originating from a bottom ($b$)  
 quark jet. 
 We                                   
 extend this data selection 
 by including also events with three jets and events  
 with ``forward'' electrons detected at pseudorapidities~\cite{foot1}
 $|\eta|>1.5$.
We also now accept  
the small contribution originating from misreconstructed
$ZH$, in which only one lepton from the $Z$ is identified.
 In addition we use a more inclusive                                
 trigger selection in the muon channel, increasing                             
 the detection efficiency from approximately 70\% to                            
 100\%~\cite{lellouch}, we improve the $b$-jet identification using a                             
 neural network algorithm~\cite{nn-btag},
and we enhance the signal                        
 to background discrimination using a neural network                           
 for the $W+2$ jet events. 
Overall, the improvements in analysis techniques have led to an
increase of about 40\% in the sensitivity (for an equivalent luminosity)
to a Higgs boson with mass 115~GeV,
with respect to our previous analysis~\cite{wh-plb}.

For the $e$ channel, the $W$ + jets candidate events
are collected, with $\approx$ 90\% efficiency, by triggers that require at
least one electromagnetic (EM) object in the calorimeter. In
the $\mu$ channel, $\approx$ 90\% of the
 candidates are  collected by
         triggers requiring a single muon or a muon plus a jet, while the
         remaining ~10\% of events are collected by other triggers, for 
 a  total trigger efficiency of $\approx$ 100\%, as estimated in data~\cite{lellouch}.
%

The event selection  requires one
lepton candidate with $p_T > 15 $ GeV,
\MET$\rm{}$  $ > 20 $ GeV (\MET$\rm{}$  $ > $25 GeV for events with a forward electron),
and exactly two jets with
$p_T > 25$ and $20$ GeV, and
$|\eta| < 2.5$,
or exactly three jets with
$p_T > 25, 20$ and $20$ GeV, and  $|\eta| < 2.5$.
We also require the scalar sum of the $p_T$ of the jets to be  $ > 60$~GeV,
the
$W$ transverse mass $M^T_W$
reconstructed from the \MET$\rm{}$  and the lepton $p_T$
to be greater than
40~GeV$-$0.5$\times$\MET$\rm{}$ to reject multijet background,
and~the~primary~interaction vertex
%
to take
place within the longitudinal acceptance of the vertex detector.
Jets are reconstructed using
 a midpoint cone algorithm~\cite{blazey} with a radius
of 0.5.
%
The  \MET$\rm{}$  is calculated from energies  in calorimeter
cells
and
corrected  for the $p_T$ of identified muons.
All energy corrections applied
to electrons or
jets are also propagated to the \MET .

A central (forward) electron is required to
have  $| \eta | <  1.1$ ($1.5 < | \eta | <  2.5$).
To reject fake electrons originating mostly from instrumental
effects (track-photon overlap), the electron candidates
must
satisfy two sets of identification
(``loose'' and ``tight'') criteria~\cite{wh-plb}.
The efficiencies of these
requirements
are determined from a pure sample of
 $Z \rightarrow e^+e^-$ events.
%
The differential multijet  background for every relevant distribution
is then estimated
from the loose and tight
lepton samples~\cite{ttbar-prd,wh-plb}.
The same statistical method is used for muons, but with different
loose/tight definitions.
Muons are reconstructed using information from
the outer muon detector and the central tracker,
and must have $| \eta | <  2.0$.
To reject muons originating from semi-leptonic decays of heavy-flavor
hadrons, 
we exploit the fact
that they
have lower $p_T$  than those originating from $W$
decay, and are generally not isolated
because of accompanying jet fragments.
The loose isolation criterion is thus defined by specifying a spatial separation
between a muon and the closest jet in the
 $\eta$--$\varphi$ plane of 
$\Delta R =\sqrt{(\Delta \eta)^2 + (\Delta \varphi)^2} > 0.5$,
where $\varphi$ is the azimuthal angle.
Tighter isolation is defined by requiring little tracking
and calorimetric activity
around the muon~track.


The dominant backgrounds to {\it WH} production are from $W+$heavy flavor jets
production,
top quark pair production ($t\bar{t}$), and
single top quark production.
Signal ($WH$ and $ZH$)
and diboson processes
($WW,~WZ,~ZZ$)
are simulated using
the  {\sc pythia} \cite{emu-pythia}  event generator,
and  CTEQ6L~\cite{emu-CTEQ} leading-order
parton distribution functions.
``$W$+jets'' events refer to
$W$ bosons produced in association with light-flavor jets (originating
from $u$, $d$, $s$~quarks or gluons) 
or  charm jets (originating from $c$~quarks),
and constitute the dominant background before $b$-jet identification.
$Wc\bar{c}$ and $Wb\bar{b}$ 
are simulated individually and associated as 
``$Wb\bar{b}$''                                    
 for purposes of accounting.
These $W$ boson processes 
are generated
 with {\sc alpgen}~\cite{emu-ALPGEN} interfaced
to {\sc pythia}\ for showering and fragmentation, since {\sc alpgen}\
provides a more complete simulation of processes with high jet multiplicities
than {\sc pythia}.
The  
\ttbar\ and $Z$+jets,
events are  also generated using {\sc alpgen}/{\sc pythia}.
The production of single top quarks 
is simulated with {\sc comphep} ~\cite{emu-COMPHEP}.

%
%
The simulated backgrounds are normalized
 to their respective NLO theoretical cross sections,
 with the exception of the $W+$ jets and $W+$ heavy-flavor samples,
which are normalized to data after subtraction of all the other backgrounds,
before $b$-jet identification. 
All generated events are processed
through the D0 detector simulation based on  {\sc geant} \cite{geant}.
Data collected with a random bunch crossing trigger
are overlaid on the simulated events to model the occupancy
of the detector which is dependent on the instantaneous
 luminosity.
The resulting events are then passed through the reconstruction software.
Finally, corrections are applied
to account for the trigger efficiency and for
residual discrepancies between the data and the simulation.

We use a neural network $b$-tagging (NN$_{b}$) algorithm~\cite{nn-btag} 
to identify heavy-flavor jets.
Its requirements are optimized 
for the best sensitivity to the Higgs boson signal.
For each  jet multiplicity, we form two statistically independent samples, one
(2~$b$-tag) with two $b$-tagged jets using a 
loose NN$_{b}$ criterion resulting in a
$b$-jet efficiency of 59\% and a light-jet tagging (mistag) probability
of 1.7\%, and a
second (1~$b$-tag)  with exactly one $b$-tagged jet using a tighter
NN$_{b}$ criterion (48\% efficiency and 0.5\% mistag
probability).
All efficiencies are determined for jets satisfying minimum 
requirements
in terms of track quality and multiplicity (``taggable jets''), 
which constitute $\approx$ 80\% of all jets.
%
%
%
%
In the simulations, the $b$-tagged  jets are weighted to
reproduce the tagging rate measured in data
samples.

\begin{figure}[b]
\psfig{
figure =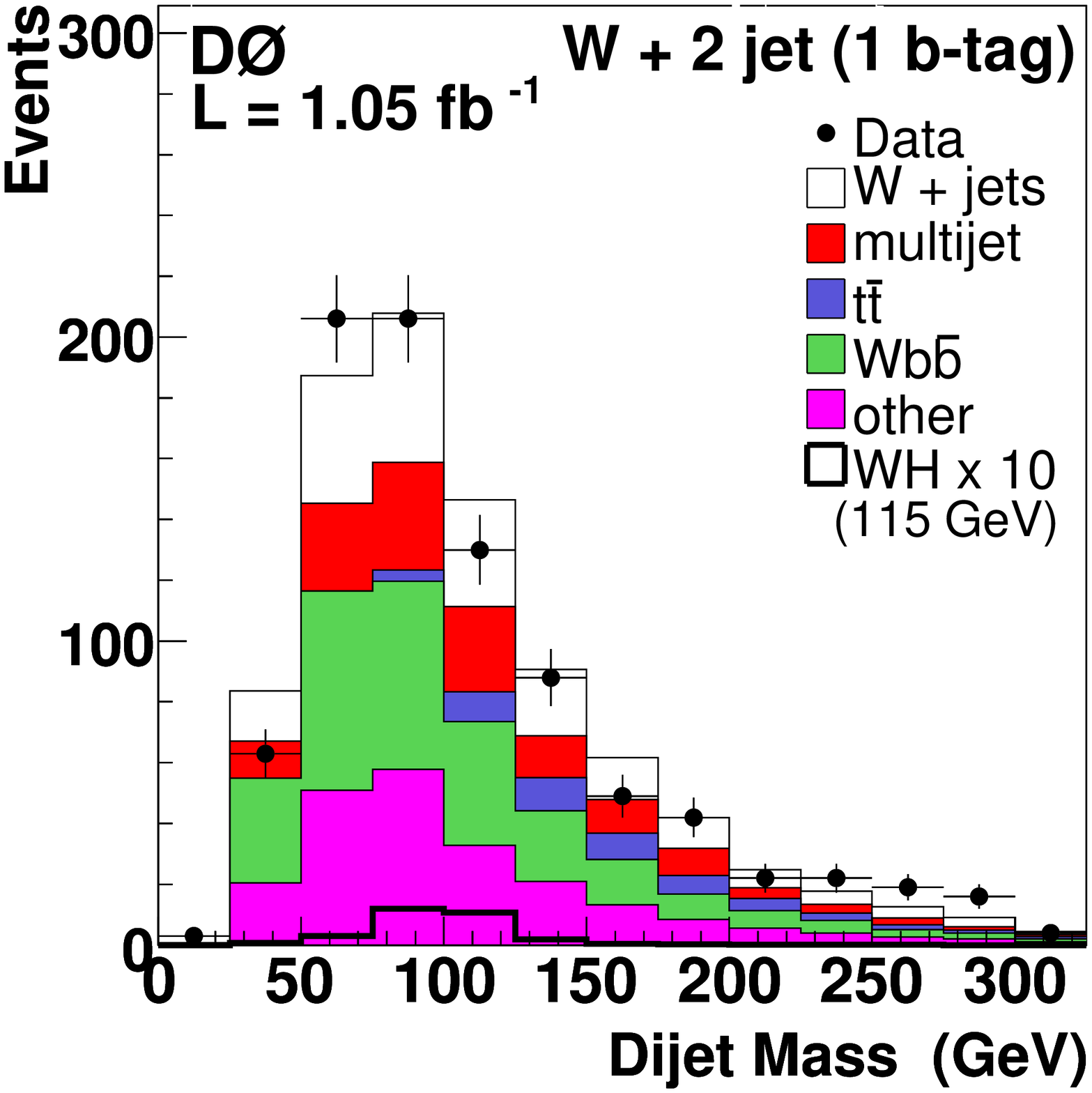,height=4.2cm}
\psfig{
figure =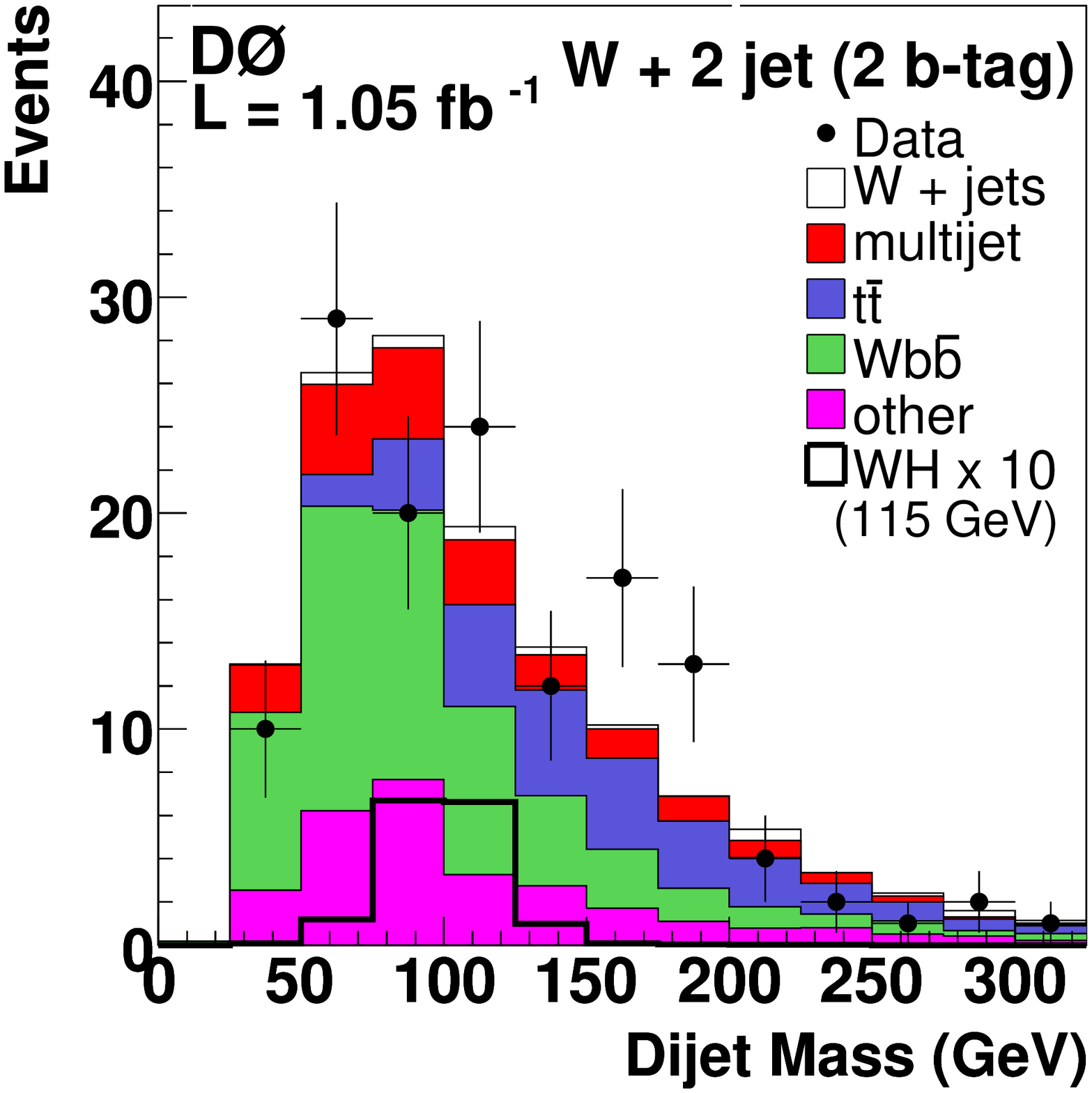,height=4.2cm}\\
\psfig{
figure =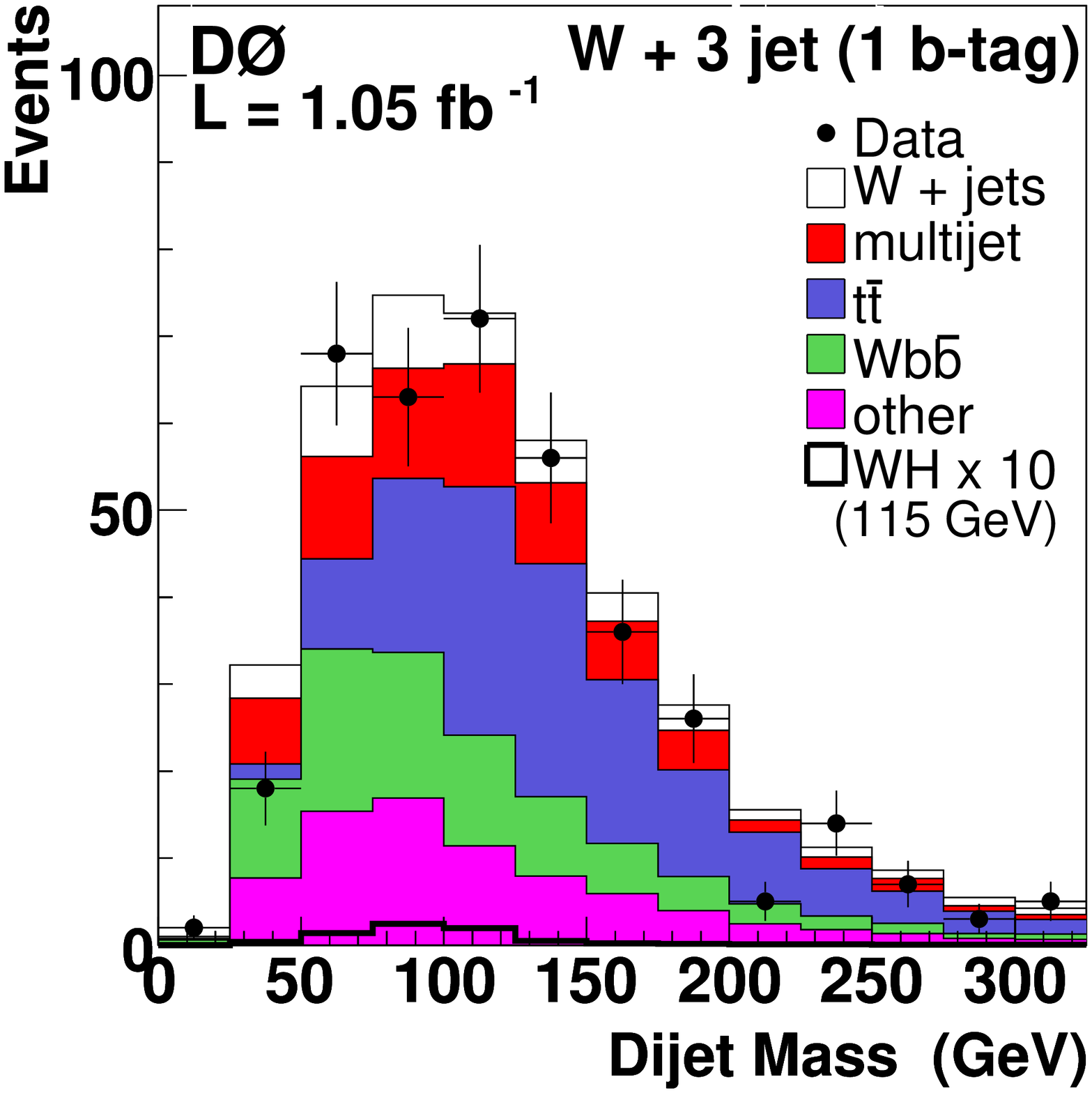,height=4.2cm}
\psfig{
figure =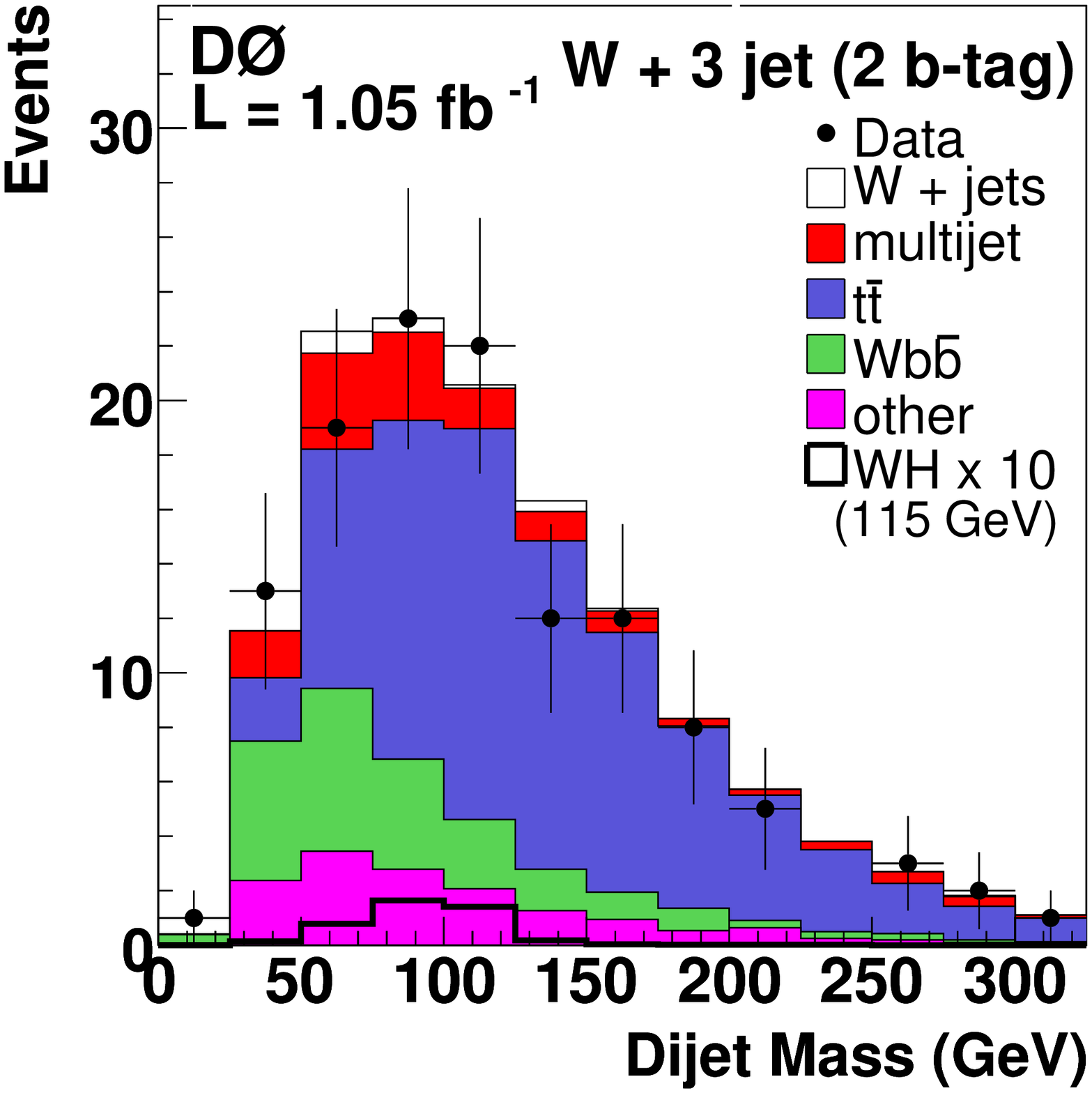,height=4.2cm}\\
\psfig{
figure =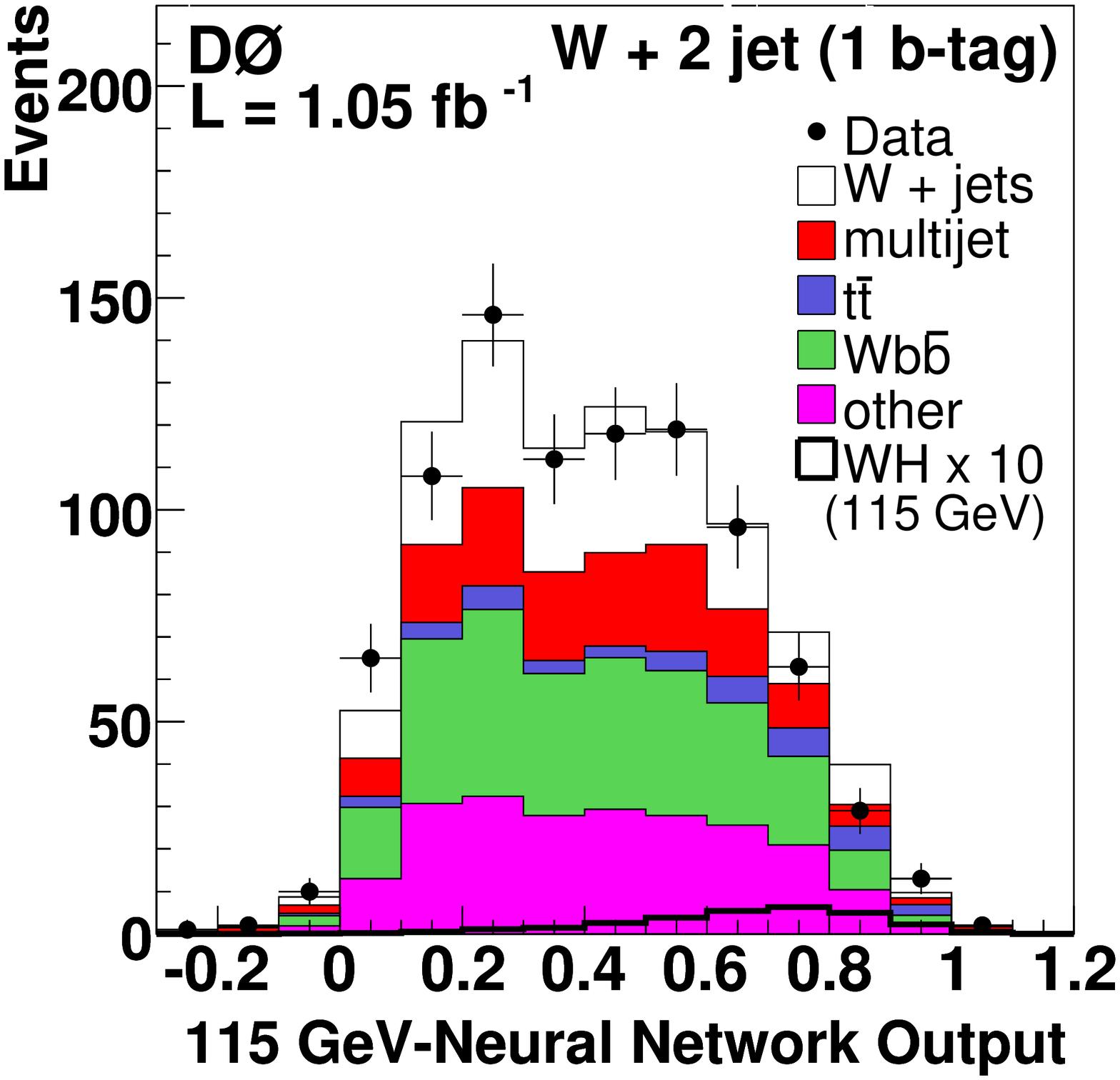,height=4.2cm}
\psfig{
figure =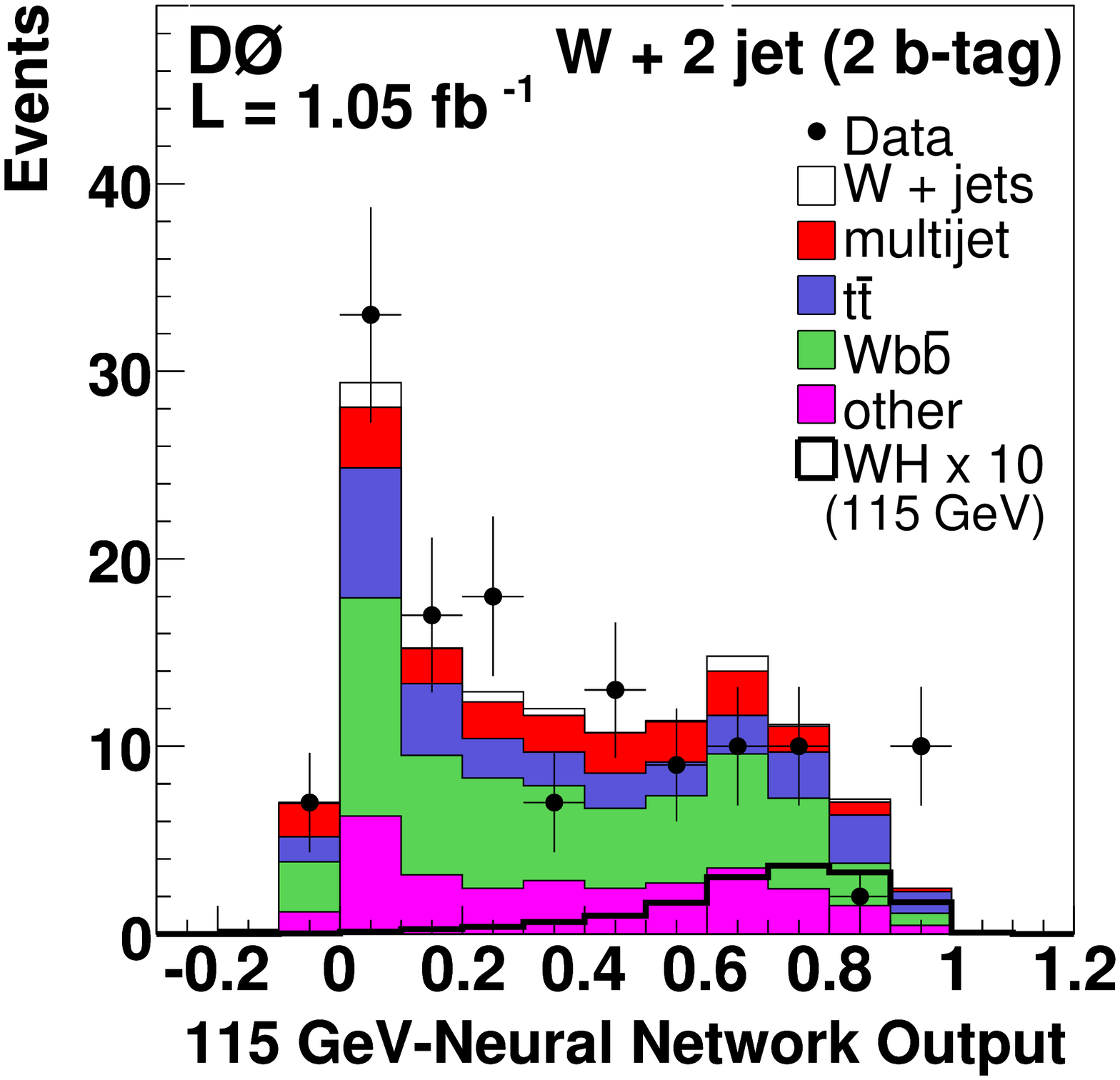,height=4.2cm}\\
 \begin{picture}(0,0)(0,0)
  \put (-25,300){\scriptsize {\bf (a)}}
  \put ( 95,300){\scriptsize {\bf (b)}}
  \put (-25,180){\scriptsize {\bf (c)}}
  \put ( 95,180){\scriptsize {\bf (d)}}
  \put (-25, 60){\scriptsize {\bf (e)}}
  \put ( 95, 60){\scriptsize {\bf (f)}}
 \end{picture}
\caption{\small
Dijet mass distributions for the $W+2$ jet,
$W+3$ jet
1~$b$-tag events (a,c) and 2~$b$-tag (b,d) events.
The data are
compared to the background prediction.
The distributions in the neural network discriminant for $W+2$ jet
1~$b$-tag and 2~$b$-tag events are shown in (e,f), respectively.
The expectation from {\it WH}(x10) production for $m_H = 115$ GeV is overlaid
 (color online).
}
\label{emu-two-tags}
\end{figure}

Using these selection criteria,
we observe 885 (385) in the 1~$b$-tag  $W+2$ jet ($W+3$ jet) samples
and 136 (122) events in the corresponding 2~$b$-tag sample.
Distributions of the dijet invariant mass, using the two jets of highest $p_T$,
 in $W+2$ jet and $W+3$ jet events  are  shown for the
1~$b$-tag and 2~$b$-tag samples in Fig.~\ref{emu-two-tags}(a--d).
The data are well described by the sum of the simulated
SM processes and 
multijet background.
The expected contributions from a  Higgs boson with $m_H=115$ GeV
are also shown.
The expected event yields for the backgrounds and a Higgs boson                         
 with $m_H=115$ GeV 
are compared to the observed number of events in 
Table~\ref{emu-tab:table3}.
\begin{table}[t]
\begin{center}
\caption{\label{emu-tab:table3} {
Summary of event yields for  the
$\ell$ ($e$ and  $\mu$) + $b$-tagged jets +  \MET$\rm{}$  final state.
Events in data are compared with the
expected number of 1~$b$-tag and 2~$b$-tag events in the   $W + 2$ and $W+3$ jet samples,
in simulated samples of diboson (labelled ``$WZ$'' in the table),
$W/Z + b \bar{b}$ or $c \bar{c}$ (``$W b\bar{b}$''), $W$/$Z+$ light quark jets 
(``$W$+jets''), 
top quark (``$t \bar{t}$'' and ``single $t$'') production,
and  multijet background (``m-jet'') determined
from data (see text). 
The  {\it WH} expectation 
 is given for $m_H=115$ GeV, and not included in the ``Total'' SM expectation.
}
}
\begin{tabular}{lrclrclrclrcl}
\hline
\hline
                &  $W $&+&$$ 2 jet   &  $W $&+&$ 2$ jet   &  $W $&+&$ 3$ jet   &  $W $&+&$ 3$ jet   \\
     & \multicolumn{3}{c} { 1$b$-tag }& \multicolumn{3}{c} {  2 $b$-tag  }
                                            & \multicolumn{3}{c} { 1 $b$-tag } & \multicolumn{3}{c} { 2 $b$-tag }\\
\hline
 {\it WH}      &   2.8 &$\pm$& 0.3 &    1.5 &$\pm$& 0.2  &   0.7&$\pm$&0.1   &  0.4 &$\pm$& 0.1 \\
\hline
$WZ$       &   34.5 &$\pm$&3.7 &    5.3 &$\pm$& 0.6  &    9.1&$\pm$& 1.0 &   1.7 &$\pm$& 0.2 \\
$Wb\bar{b}$&  268   &$\pm$& 67 &   54   &$\pm$& 14   &   87&$\pm$&22     &   22.7  &$\pm$&  5.7 \\
$W$+jets   &  347   &$\pm$& 87 &   14.0 &$\pm$&  4.4 &   96&$\pm$&24     &   8.5 &$\pm$&  2.7 \\
$t\bar{t}$ &   95   &$\pm$& 17 &   37.4 &$\pm$&  7.0 &  156&$\pm$&29     &   81  &$\pm$& 15 \\
single $t$ &   49.4 &$\pm$&9.0 &   12.4 &$\pm$&  2.3 &  15.7&$\pm$&2.9   &   6.7 &$\pm$&  1.2 \\
m-jet      &  104   &$\pm$& 29 &    8.9 &$\pm$&  2.1 &   54&$\pm$&15     &   8.7 &$\pm$&  2.1 \\
\hline
Total      &  896   &$\pm$&177 &  132   &$\pm$& 27    &  418&$\pm$&76    &  129  &$\pm$& 24   \\
Data       &    885          &&&    136            &&&    385            &&&   122   &&         \\
\hline
\hline
\end{tabular}
\end{center}
\end{table}

Although the dijet invariant mass is a powerful variable for
separating a Higgs boson signal from background~\cite{wh-plb}, the sensitivity of the
analysis is enhanced through the use of multivariate techniques:
in $W$+2 jet events,
a neural network is trained
on simulated signal and $Wb\bar{b}$ events, using seven
kinematic variables: $p_T$ of the highest and second-highest $p_T$~jet,
$\Delta R$(jet$_1$,jet$_2$), $\Delta \varphi$(jet$_1$,jet$_2$),
$p_T$(dijet system), dijet invariant
mass, and $p_{T}$($W$ boson candidate). The training is performed
for every simulated Higgs signal (different test masses), and
separately for $e$, $\mu$, 1~$b$-tag and 2~$b$-tag events.
The resulting neural networks are then applied to  $W+2$ jet data
and to the background and simulated signal samples. In the final
limit-setting procedure, the distributions of the neural network
discriminant
  corresponding to 
a specific Higgs boson test mass  are used for analyzing the $W+2$ jet events.
The improvement in sensitivity over just using
the dijet invariant mass is about 15\% at $m_H = 115$ GeV.
The resulting neural network  discriminants are shown in
Fig.~\ref{emu-two-tags}(e,f).
For the $W+3$ jet samples, whose dominant background is \ttbar , 
the limits are determined directly from the
dijet mass distributions.

Systematic uncertainties
on  efficiencies
and from the propagation of
other systematics (e.g. energy calibration and detector response)
are: 
 (3--5)\% for trigger efficiency;
 (4--5)\%  for lepton identification efficiency;
 6\% for jet identification efficiency and jet resolution; 
 5\% from the modeling of the jet multiplicity spectrum;
3\%  due to
  the uncertainty in the jet energy calibration;
 2--10\%
 due to the uncertainty in modeling $W+$jets, determined by comparing data and expectation
 before $b$-tagging and before reweighting the $W+$ jet samples to match the data 
(the effect of this uncertainty on the shape
 of the neural network discriminant is also taken into
 account);
 3\% for jet taggability; and
2\% uncertainty for $b$-tagging efficiency.
 For light quark jets, the uncertainty on the mistag rate is
15\%.
The multijet background, determined from data, has an uncertainty
of 18--38\%.
The systematic uncertainty on the theoretical cross section for the simulated backgrounds
is 6--20\%, depending on the process.
The uncertainty on the luminosity is 6\%~\cite{emu-lumi}.

We use the CLs method~\cite{emu-junkLim,emu-wadeLim} 
to assess the compatibility of data                                    
with the presence of a Higgs signal. In the absence of any                                       
significant enhancement, we obtain upper limits 
%
on {\it WH} production,
using the neural network output (dijet invariant mass of the $b \bar{b}$ system)
for the $W+2$ jet ($W +$ 3 jet) sample
as the final discriminating variable.
The 1~$b$-tag and 2~$b$-tag, and the $e$ and $\mu$ channels, are treated
separately,
giving a total of eight analyses, which are then combined~\cite{wh-plb}.
  We incorporate sytematic uncertainties on signal and
background expectations using Gaussian sampling and  include correlations among the
uncertainties across the analysis channels. We reduce the impact of sytematic uncertainties
 using the  profile likelihood technique~\cite{emu-wadeLim}.
%
%
\begin{figure}[b]
 \centerline{
 \psfig{
figure=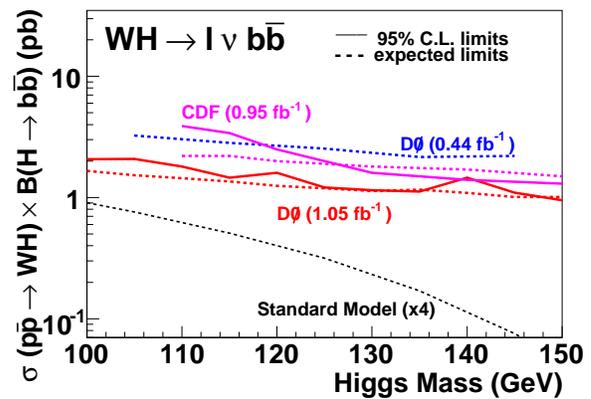,height=6.5cm}}
\caption{ \label{emu-limits_plot}
95\% C.L.  cross section upper limit (and corresponding expected limit)
 on
$\sigma(p\bar{p} \rightarrow WH) \times B(H \rightarrow b \bar{b})$ 
 vs. Higgs boson mass,
compared to
the SM expectation and to 
the expected limit from our previous
analysis~\cite{wh-plb}.
Recent CDF  results~\cite{emu-CDF-wh-1fb}
 are also shown.
Solid (dashed) lines represent observed (expected) limits.
The contribution of {\it ZH} reconstructed in the same final state is
taken into account in the {\it WH} signal when deriving the limits,
assuming the SM ratio of {\it ZH/WH} cross sections.
}
\end{figure}

%
\begin{table}[t]
\begin{center}
\caption{\label{emu-limits-wh}{ Observed and  expected  95\% C.L. upper limits
on the cross section times branching fraction
($\sigma \times B$) in pb, where $B=B$($H \rightarrow b \bar{b}$),
for different Higgs boson mass values;
the corresponding ratios to the predicted SM cross section
are also given.}}
\begin{scriptsize}
\begin{tabular}{lccccccccccc}
  \hline
  \hline
$m_H$ [GeV]            &    100&    105&    110&    115&    120&    125&    130&    135&    140&    145&    150 \\
  \hline
exp.$\sigma \times B$  &   1.66&   1.53&   1.44&   1.36&   1.25&   1.19&   1.13&   1.16&   1.09&   1.01&   1.01 \\
obs.$\sigma \times B$  &   2.07&   2.08&   1.80&   1.46&   1.54&   1.21&   1.15&   1.12&   1.46&   1.10&   0.95 \\
  \hline
exp. ratio             &    7.3&    8.0&    9.2&   10.7&   12.3&   15.1&   19.1&   27.3&   37.4&   53.5&   90.2 \\
obs. ratio             &    9.1&   11.0&   11.5&   11.4&   15.1&   15.3&   19.5&   26.4&   50.1&   58.2&   83.9 \\
  \hline
  \hline
\end{tabular}
\end{scriptsize}
\end{center}
\end{table}

The combined upper limits obtained at the 95\% C.L.
on $\sigma(p\bar{p} \rightarrow WH) \times B(H \rightarrow b \bar{b})$
 are displayed in Fig.~\ref{emu-limits_plot}
and given in Table~\ref{emu-limits-wh},
together with
the ratios of these limits to the predicted SM cross section.
For this analysis, all deviations between observed and expected limits are less than
1.5 standard deviations.
At $m_H = 115$ GeV, the observed (expected) limits are 1.5 (1.4) pb, or a factor of 11.4
(10.7) times
higher than the SM prediction.
Our new limits are displayed in Fig.~\ref{emu-limits_plot}
and compared
to the expected limit from our                                  
previous analysis~\cite{wh-plb}.  The improvement in sensitivity is                                  
significant, and our expected limits  scale approximately                                 
inversely with luminosity compared to our previous result. These limits                              
are the most stringent to date in this process at a hadron collider.                                 
%
%

In summary,
we have presented 95\% C.L. upper limits on the product of $WH
\rightarrow \ell \nu b \bar{b}$  production cross
section and branching fraction for $H \rightarrow b\bar{b}$.
These range between 2.1 and 1.0 pb for
$100 <  m_H < $ 150 GeV, while the corresponding
SM predictions range from 0.23 to 0.01 pb.
The sensitivity should increase significantly in
the near future with the continuing accumulation of luminosity
from  the Tevatron and improvement in analysis techniques.
A significant sensitivity gain has already been
achieved by combining these data, with other Higgs boson searches
done by the CDF and D0 collaborations~\cite{full-combo}.

%
%


%
We thank the staffs at Fermilab and collaborating institutions,
and acknowledge support from the
DOE and NSF (USA);
CEA and CNRS/IN2P3 (France);
FASI, Rosatom and RFBR (Russia);
CNPq, FAPERJ, FAPESP and FUNDUNESP (Brazil);
DAE and DST (India);
Colciencias (Colombia);
CONACyT (Mexico);
KRF and KOSEF (Korea);
CONICET and UBACyT (Argentina);
FOM (The Netherlands);
STFC (United Kingdom);
MSMT and GACR (Czech Republic);
CRC Program, CFI, NSERC and WestGrid Project (Canada);
BMBF and DFG (Germany);
SFI (Ireland);
the Swedish Research Council (Sweden);
CAS and CNSF (China);
the Alexander von Humboldt Foundation (Germany).


\begin{thebibliography}{99}
\bibitem{foot1}
The pseudorapidity is given in terms
of the polar angle $\theta$ as $\eta \equiv - \ln(\tan{\frac{\theta}{2}})$,
where $\theta$ is defined relative to the center of the detector.
\end{thebibliography}
\end{document}